\documentclass{JHEP3}

\newcommand{\be}{\begin{equation}}
\newcommand{\ee}{\end{equation}}
\newcommand{\bea}{\begin{eqnarray}}
\newcommand{\eea}{\end{eqnarray}}
 \newcommand{\IC}{\mathbb{C}}

\newcommand{\non}{\nonumber \\}

\def\IZ{\relax\ifmmode\hbox{Z\kern-.4em Z}\else{Z\kern-.4em Z}\fi}

 \def\bh{{\bar h}}

\def\bj{{\bar j}} \def\bk{{\bar k}}

 \def\cn{{\cal N}}
\def\cl{{\cal L}} \def\co{{\cal O}}

\def\tr{{\rm tr}}

\def\cn{{\cal N}} \def\cM{{\cal M}} \def\Mc{{\cal M}_c}
\newcommand{\sbsection}[1]{\vspace{.5cm} \noindent {\it #1}}

\preprint{{\tt hep-th/0205141}}

\title{On Conformal Deformations}

\author{Barak Kol
\\
School of Natural Sciences \\ Institute for Advanced Study \\
Einstein Drive, Princeton NJ 08540,
USA\\
\email{barak@sns.ias.edu} }

\abstract{For a conformal theory it is natural to seek the
conformal moduli space, $\Mc$ to which it belongs, generated by
the exactly marginal deformations. By now we should have the tools
to determine $\Mc$ in the presence of enough supersymmetry. Here
it is shown that its dimension is determined in terms of a certain
index. Moreover, the D-term of the global group is an obstruction
for deformation, in presence of a certain amount of preserved
supersymmetry. As an example we find that the deformations of the
membrane (3d) field theory, under certain conditions, are in
${\bf 35}/SL(4,\IC)$. Other properties including the local
geometry of $\Mc$ are discussed.}

\begin{document}

\begin{flushright} \begin{tabular}{l} {\it To my daughter,} \\ {\it
Inbal.}
\end{tabular} \end{flushright}


\section{Introduction}

Given a field theory one of the most basic properties that one
would like to investigate is the vacuum, or in general the moduli
space of vacua. Indeed, in the last decade we have learned a lot
about moduli spaces of vacua for supersymmetric (susy) field
theories. Given a conformal field theory, an equally natural
question is to study the moduli space of conformal theories on
which it lies, namely the space of parameters for which the theory
is exactly conformal. Since both the moduli space of vacua and the
conformal moduli space may be interesting for the same field
theory I denote the first conventionally by $\cM$ and the latter
which is the subject of this paper by $\Mc$.

The moduli space of vacua $\cM$ is generically expected to be
trivial (a point or none) and only in the presence of a certain
amount of supersymmetry it is generic to have a non-trivial
manifold. Similarly $\Mc$ is expected to be trivial in the
absence of any symmetry such as supersymmetry. Therefore our
first task, which is the main subject of this paper, is to
determine the dimension of $\Mc$.

Looking beyond the dimension we would like to know what is the
local geometry of $\Mc$ (once we define its metric) - the relevant
analogue of metrics with reduced holonomy. Moreover, we may
investigate global issues of $\Mc$ - the location of
singularities, (non-)compactness, and possibly a non-trivial
topology.

\sbsection{Summary of results}

We begin by discussing some background in section
\ref{background}. Leigh and Strassler \cite{LS} (see also
references therein) described a general mechanism to find the
dimension of $\Mc$ for 4 dimensional $\cn=1$ theories. Despite
its success, this formulation raises certain concerns of gauge
invariance and scheme dependence. Moreover, it uses cleverly
chosen operators with some ad-hoc symmetry properties, breaking
the covariance of the problem under the global symmetry, and it
is not clear that {\it all} exactly marginal operators are found
in this way.

In order to gain further insight into the issue we use the AdS-CFT
correspondence \cite{AdSCFT} to discuss the analogous concept in
gravity, namely the space of vacua which preserves the conformal
isometries, or equivalently the AdS factor. In this way $\Mc$ is
understood to be related, when a duality exists, to an $\cM$ of
the (gauged) gravity, a concept that deserves further study, much
in same way that the $\Mc$ of the worldsheet theory is the $\cM$
of string theory. However, the original formulation of \cite{LS}
does not translate directly to supergravity. By understanding the
supergravity point of view we can abstract properties of $\Mc$
which are independent of the existence of a gravity dual, and
refine the formulation of \cite{LS} so that the issues raised
above are clarified.

In this paper we present the following results
\begin{itemize}
\item The conformal index
\item A D-term obstruction
\item Example: the membrane field theory
\end{itemize}
which we now discuss one by one.

Like many other deformation problems the (virtual) dimension of
$\Mc$ can be formulated in terms of an index (section
\ref{section_index}). The relevant operator here is the
supersymmetry variation, and in that respect the conformal index
can be considered to be a special example of the Witten index
\cite{WittenIndex}. This point of view is certainly standard for
2d field theories, and is hidden in the 4d counting argument of
\cite{LS}, but I am not aware of previous discussions of it in the
literature.

An index formulation is useful only after a practical method of
computing it is found. Since the index is nothing but the
difference of the number of zero modes and the number of
obstructions, and moreover the number of zero modes, or
supersymmetric marginal deformations, is usually readily
determined, we concentrate on the description of the obstruction.
In section \ref{section_obstruction} I show that the D-term for
the {\it global group} is such an obstruction for field theories
with 8 super-conformal charges or more (such as 4d $\cn=1$). This
is not surprising if we remember that the global group becomes the
local gauge group in the gravity dual, and in some sense on a
general $\Mc$. Moreover, since we may think of the field theory
parameters as the VEVs of a background chiral field, we expect
$\Mc$ to have a complex structure, and since we wish to divide
the space of couplings by the global group, we need to take the
D-term constraint in order to achieve a holomorphic quotient.
When comparing to \cite{LS} the D-term should be considered to be
a refinement, or a more precise replacement, for the gamma
function constraints used there (the replacement should take
place in the NSVZ formula \cite{NSVZ} as well).

The case of the field theory of the membrane is discussed in
section \ref{example}. This is a 2+1 conformal field theory with
an $SO(8)$ global symmetry. It depends on a choice of an A-D-E
group, which is here taken to be $A_N$ for some large enough $N$
to conform with supergravity. Using the methods described above we
find that for deformations that preserve 8 super-conformal
charges $\Mc \simeq {\bf 35}/SL(4,\IC)$ locally (up to a possible
finite $N$ effect). This is  a $20_\IC$ dimensional manifold,
where ${\bf 35}$ is the fourth rank symmetric tensor of $SU(4)$.

\sbsection{Discussion and open questions}

In this paper we discuss the dimension of $\Mc$ and we do not
touch much onto the local geometry nor onto global issues. The
study of the local geometry can be initiated through the gravity
dual by considering the local geometry of $\cM$ for 5d gauged
supergravity (minimal susy). Quite a lot is known about these
theories \footnote{See \cite{Ceresole-DallAgata} for the
state-of-the-art.}: the scalars in hypers live in a quaternionic
manifold, while the scalars in vectors and tensors live in a
``very special geometry'', and the metric is related as usual to
the kinetic terms of the moduli. The vacuum manifold is
determined by ${\cal V}=0$ where ${\cal V}$ is the scalar
potential. However, the structure of the resulting sub-manifold
is not well-understood, but hopefully it is within reach. In
addition it would be interesting to determine which field theory
couplings types (in 4d $\cn=1$: various superpotential couplings,
gauge couplings) correspond to which supergravity multiplets
types (in 5d: hyper, vector or tensor).

Hopefully at some point we will have a complete description of
some $\Mc$'s (similar to the Seiberg-Witten solution of $\cM$ for
4d $\cn=4$), and then we will be able to address some issues
about the region of $\Mc$ away from the neighborhood of the
origin. It would be interesting to know whether $\Mc$ has any
compact factors, and whether there are singularities (we will see
that the origin is generally potentially singular because of the
quotient structure).

The determination of the dimension of $\Mc$ in this paper is
incomplete in some ways. In particular I would like to note here
that for theories which may allow susy deformations with fewer
than 8 super-conformal charges (such as the membrane theory) a
larger $\Mc$ which contains the one found here may exist
(currently under study).

A related topic worth pointing out is that
spontaneous\footnote{Spontaneous from the supergravity point of
view.} partial supersymmetry breaking is readily achieved on
$\Mc$. For example 4d $\cn=4$ can be deformed to $\cn=1$ ($1/4$
susy) and similarly for the membrane theory.

\section{Background: field theory and gauged supergravity}
\label{background}

A $d$ dimensional conformal field theory (CFT) is by definition a
Poincar\'{e} invariant, scale invariant theory which
together with the special conformal transformations is invariant under the full
$SO(d,2)$ conformal group.

A CFT is conventionally defined as a set of operators together
with their correlation functions. Given such a theory, its
deformations are one to one with the set of operators, and we may
think of deforming with an operator $\co_d$ as adding it to the
Lagrangian \footnote{When a Lagrangian description is available.}
\be \cl \to \cl+ h\, \co_d \ee
 where $h$ is the expansion parameter. The effect on an arbitrary correlation
 function
 \be <\co_1(x_1)\, \co_2(x_2)>={\int {\cal D}\phi\, \co_1(x_1)\,
 \co_2(x_2)\, \exp (i\int d^d x\, \cl) \over \int {\cal D}\phi\, \exp (i\int d^d x\,
 \cl)} \ee
(expressed in terms of a functional integral over all fields
$\phi$) 
 is
  \bea <\co_1(x_1)\, \co_2(x_2)>  ~\to ~~~~~~~~~~~~~~~~~~~~~~~~~~~~~~~~~~~~~~~~~~~ \non
  <\co_1(x_1)\, \co_2(x_2)>_h = {\int {\cal D}\phi\,
\co_1(x_1)\, \co_2(x_2)\, \exp [i\int d^d x\, (\cl+h\, \co_d)] \over \int {\cal D}\phi\, \exp [i\int d^d x\, (\cl+h\, \co_d)]} = \non
 = <\co_1(x_1),\co_2(x_2)>_0 + h \int d^d x <\co_1(x_1)\, \co_2(x_2)\,
 \co_d(x)> + O(h^2)
 \eea
and the triple correlation function is understood to be averaged
over all orderings. This series is usually referred to as {\it conformal
perturbation theory}.

We would like to study conformal deformations, namely operators
$\co_d$ such that the theory remains conformal to all orders of
$h$. To first order this is equivalent to the dimension of $\co_d$
being $d$, and such operators are called {\it marginal}, while the
ones which are conformal to all orders are called {\it exactly
marginal}. An operator that is both marginal and preserves the
supersymmetry of the theory will be called here
``super-marginal''.

\sbsection{Leigh - Strassler}

Leigh and Strassler \cite{LS} (see also references therein)
described how the existence of a non-trivial $\Mc$ may be deduced
for some 4d $\cn=1$ theories. We shall describe their results
here in their language, and later we will see that some
refinements are required. Couplings in 4d $\cn=1$ may be divided
into (complex) gauge couplings and superpotential parameters. A
set of couplings $h_i$ is exactly marginal if and only if all
their beta functions $\beta_i$ vanish (to all orders in $h_i$).
The starting point are the exact $\cn=1$ formulas for the beta
functions in terms of the gamma functions of the charged fields.
For the gauge coupling, $g$, it is the NSVZ formula
 \be \beta_g \sim f(g)\, [\beta_0 + \gamma] \ee
in terms of $\beta_0$, the 1-loop beta function, and the gamma
functions; And for the beta function of a superpotential parameter
$h$ ($\delta W = h\, \co$) it is
 \be \beta_h \sim h\, [\beta_0 + \gamma] \ee
where $\beta_0=-\Delta_W+\Delta_\co$, $\Delta_W=d-1$ is the
dimension of the superpotential in $d$ dimensions, and
$\gamma=\gamma(\co)$. From this linear dependence it is deduced
that for marginal operators ($\beta_0=0$) it is enough to set to
zero all the gamma functions. If due to some symmetries the
number of gamma functions is fewer than the number of
(super-marginal) couplings, a non-trivial $\Mc$ will exist
(generically) simply from counting unknowns and constraint
equations.

One concern presents itself immediately, namely, for gauge
theories the gamma functions used here are related to 2-point
correlation functions $<\phi(x_1)\, \phi(x_2)>$ which are not
gauge invariant. This can be circumvented either by considering
2-point correlation functions of gauge invariant composites of
$\phi$ such as $<\tr (\phi^2)(x_1)\,\tr (\phi^2)(x_2)>$ and
trying to read $\gamma$ off the result, or by defining $\gamma'$
by inserting a Wilson line into the correlator along some
arbitrary line $<\phi(x_1) \int_{x1}^{x2} A_\mu(x)\, dx^\mu \,
\phi(x_2)>$, and then arguing that at weak coupling $\gamma$ and
$\gamma'$ become the same, but neither method is quite
satisfactory. Another concern raised in the past was the scheme
dependence of both the beta functions and the associated gamma
functions beyond 1-loop.

 \sbsection{Supergravity}

For conformal field theories which possess an AdS gravity dual
through the AdS-CFT correspondence \cite{AdSCFT}, let us explore
how the issue of $\Mc$ translates into gravity. Suppose the dual
is $AdS_{d+1} \times X$, then we would like to deform it keeping
all the conformal isometries, which is the same as keeping the
$AdS_{d+1}$ factor intact. The most general metric ansatz is an
$AdS_{d+1}$ fibered over a deformed $X'$, with some warp function
$\rho(x), \, x \in X$. The ansatz allows to turn on any other
fields, with any $x$ dependence, as long as they are AdS scalars.

Since we currently have a working definition of string theory on
AdS only in the large radius supergravity limit, we will consider
only supergravity fields, and our results translate to the
corresponding limits of the field theory, such as 4d field theory
in the 't Hooft limit with large 't Hooft coupling, and some 3d
theories with a certain large $N$.

Finding a conformal deformation amounts now to finding a
continuous deformation from the zeroth order solution $AdS_{d+1}
\times X$ that solves the supergravity equations of motion within
this ansatz. Since we are mostly interested in supersymmetric
solutions, we will require that the solution satisfies actually
the susy variation equations on fermions (which imply the
equations of motions). The field mode in the first order (in $h$)
deformation can be directly translated from the supergravity to
field theory using the correspondence dictionary. The field theory
interpretation of the higher order modes is less clear.

\section{The conformal index}
\label{section_index}

Consider finding super-conformal deformations of a gravity
solution. In general, the (virtual) number of deformations of a
solution to a non-linear set of equations is given by the index of
the linearized equations. Let us recall why. One seeks a deformed
solution where the fields, denoted here collectively as $\phi$,
are written as a power series in a perturbation parameters $h$:
$\phi=\sum \phi^{(i)}\, h^i$. After substituting the perturbation
series in the equations one attempts to solve the equations order
by order. By assumption the zeroth order equations are satisfied,
so we go on to the linearized equations, and let us denote that
linear operator by $L$. In a diagonalized form $L \, \phi_j=L_j \,
\phi_j$  most field modes $\phi_j$ will have a non-zero eigenvalue
$L_j \ne 0$, however, we take special note of the {\it zero
modes} (or kernel) where $L_j=0$, and of the {\it obstructions}
(or cokernel) where the fields do not appear at all at linear
orders, namely these modes are outside the image of $L$. If $L$
has no zero modes then clearly there are no deformations, so we
will assume that some do exist. As we go to higher orders the
equations will be $L \, \phi^{(k)}=\dots$, where $k$ is the order
and $\dots$ denotes an expression that depends only on fields
from lower orders. For the non-zero-modes there is a unique
solution to these equations. For the zero-modes we are free to add
terms at any order, but since that amounts only to a redefinition
of the perturbation parameters, we choose not to have any such
terms beyond the first order. The obstruction equations are the
source of trouble - since they are of the form $0=\dots$ they are
a {\it constraint} on the first order deformations. In essence the
situation is like a set of non-linear equations where the zero
modes are the essential unknowns and the obstructions are the
essential constraint equations, and so the number of solutions
(or the virtual dimension of the solution space) is {\it
generically} \be \mbox{ \#(zero modes)-\#(obstructions)} \equiv
\mbox{Index}(L). \ee We have to qualify the dimension as
``virtual'' or ``generic'' since the actual dimension could be
larger if the obstructions are not independent, or it could also
be smaller when the equations are not holomorphic.

Considering the index of the supergravity {\it equations of
motion} (e.o.m.) we find that the spaces of fields and equations
are the same, so each zero mode is also an obstruction, and hence
the index is zero, and generically one cannot expect deformations
to exist (this is shown explicitly in the example of \cite{AKY}).
However, as commonly happens, supersymmetry helps. If we consider
instead of the equations of motion the {\it susy variation of
fermions}\footnote{The susy variation of the bosons also play a
role in the definition of the index.} then the fields are bosonic
modes while the constraints are fermionic, and hence the zero
modes and obstructions are not correlated and the index
generically is non-zero. From the point of view of the e.o.m.
this shows up as a degeneracy  of its obstructions.

Summing up, we see that in supergravity the number of deformations
is the index of the linearized susy variations\footnote{I thank E. Witten
 for an important discussion on this topic.}. Finding the explicit
index formula for some examples is now work in progress.
 Proceeding to field theory, since the susy
variations in supergravity map to the super-conformal charges, it
must be that the number of conformal deformations in field theory
is given by their respective index, where this time we include
all operators, not only those which are preserved in the
supergravity limit. We may call this the {\it conformal index},
and it is really a special type of the general supersymmetric
Witten index.
 \be \mbox{dim}(\Mc) = \mbox{Index} [\delta_{susy}\, ] \ee
 The zero modes are the super-marginal (bosonic) operators, for example, in
4d $\cn=1$, consider parameters in the superpotential whose
operators have protected dimensions (independent of the gauge
couplings), while the obstructions can be identified with certain
fermionic operators (or equivalently by their bosonic
super-partners), which are less obvious and in the next section I
describe some.

We can now compare with the language of \cite{LS} - their
``couplings'' must be\footnote{or at least contain} the zero
modes above, and their ``gamma function constraints'' must
be\footnotemark[\value{footnote}] the obstructions. Moreover, in
\cite{LS} the construction seems to rely on choosing
intelligently the operators so they have some high degree of
symmetry and break the global symmetry of the problem, whereas the
index language does not require that. Nevertheless, formulating
the problem in terms of an index is not useful before we find a
practical way to compute it. To that purpose we discuss the
obstructions in the next section.

\section{The D-term obstruction: $\gamma \mapsto D$}
\label{section_obstruction}

In this section we will demonstrate that for CFT's with at least 8
super-conformal charges the D-term of the {\it global group} is
an obstruction to the deformations in the sense of the previous
section.  Comparing with \cite{LS} which identified the
obstruction to be gamma functions which have problems with gauge
invariance as discussed in section \ref{background}, the D-term
can be thought to be a refinement summarized by the replacement
rule \be
 \gamma \mapsto D^I = \sum_h h^\dagger\, T^I\, h  \ee
where the sum is over all scalar fields $h$, and $T^I$ is a
generator of the global group which acts on $h$ according to its
representation. This replacement is clearly demonstrated in the
following example.

Consider 4d $\cn=4$ with gauge group $SU(N),\, N \ge 3$. It has
$10_\IC$ zero modes, $h_{ijk}$ which transform as the third rank
symmetric representation of the global $SU(3)$. Leigh-Strassler
identified two zero modes operators with special symmetry
properties and showed that they are exactly marginal.
Alternatively, one can compute the gamma functions for all of the
zero modes, and one finds $\gamma_{i \bj} \sim h_{ikl}\, \bh_{\bj
\bk \bar{l}}$ where the traceless projection should be taken on
both sides (this can be seen at weak coupling from a perturbative
evaluation, and is actually always true by \cite{AKY}). One
notices that these constraints are exactly the D-term constraint
for the $SU(3)$ representation $h_{ijk}$, and so together with a
division by the global group we get the holomorphic
quotient\footnote{Recall that the holomorphic quotient can be
defined either by imposing the D-term and dividing by the group,
or by dividing by the complexified group, keeping only closed
orbits.} \be
 \Mc(4d,\, \cn=4) \simeq {\bf 10}/SL(3,\IC) \label{mc4} \ee
\cite{AKY} where equality holds locally near the origin of $\Mc$
(the origin is the $\cn=4$ theory) and higher order corrections
are expected away from the origin when additional modes are
incorporated\footnote{The origin of ${\bf 10}/SL(3,\IC)$ can be
shown to be smooth as a complex variety - namely there are
exactly 2 independent gauge invariant coordinates with no
relations \cite{AKY}, though the metric may still be singular,
exactly like $\IC/\IZ_k$ which is smooth as a complex variety by
working with $z'=z^k$, but may have a conical metric
singularity.}. This expression reproduces the same $2_\IC$ space
as \cite{LS}, only in a more $SU(3)$ covariant way (see
\cite{Aharony:2002tp} and references therein for other recent
examples of conformal deformations).

With this example in mind, we see why the obstruction had to be
the D-term. In 4d $\cn=1$ (or other CFT's with 8 super-conformal charges)
 all parameters may be thought to be
VEV's of background chiral multiplets, and as such they should be
valued in a space with a complex structure. This property should
apply both to the original space of super-marginal operators and
to the final space of exactly marginals. Since the transition
between the two involves a division by the global group, this
division must be done holomorphically, through the use of the
D-term.

Supergravity offers another point of view on the D-term. In our
example the dual problem is to look for conformal deformations of
$AdS^5 \times S^5$ in type IIB. From a 5d $\cn=1$\footnote{Minimal susy, or 8 supercharges,
 usually called 5d $\cn=2$ in the supergravity literature.} point of view
type IIB reduced on $S^5$ is a {\it gauged} 5d supergravity, with
infinitely many KK modes, and with the $\cn=4$ multiplets
decomposed into $\cn=1$ multiplets. The crucial point is that
this supergravity has an $SU(3)$ gauge symmetry, and with this amount of supersymmetry
 vacua are expected to satisfy a D-term constraint (see
\cite{Ceresole-DallAgata} for the state-of-the-art of this
supergravity). Moreover, a detailed analysis of the susy
variation equations, shows that a certain phenomenon appears
exactly for modes which are associated with complexified gauge
transformation \cite{AKY}, as expected from a D-term. In general,
one may consider the global group to be a gauge group not only for
supergravity duals but also, is some sense, on general $\Mc$'s.

Comparing again to \cite{LS} we see that although there is
agreement in the examples they studied, the D-term constraint is
the more precise definition of the obstruction, it saves us from
choosing special symmetric operators breaking the global group and
it gives a more unified approach, which allows for generalizations
to new results, as we see in the next example.

\section{Example: the membrane field theory}
\label{example}

As an example for the use of the observations above, let us
determine $\Mc$ for the membrane field theory. By ``the membrane
field theory'' I mean the conformal field theory of $N$
coinciding membranes in M-theory (M2 branes). It is a 2+1
dimensional CFT with an $SO(8)$ global symmetry associated with
the 8 transverse directions. The theory lacks an intrinsic
definition (in particular there is no Lagrangian), and one may
indirectly define it to be the IR fixed point of a 3d $\cn=8$
(maximal) $SU(N)$ gauge theory (hence known as the ``$A_{N-1}$
theory''). I would like to show that although we know very little
about it, we {\it can} find the number of exactly marginal
deformations.

We approach the problem by considering the gravity dual which is
11d supergravity on $AdS_4 \times S^7$ with a radius that
increases with $N$, so for $N \to \infty$ we may use the 11d
supergravity limit. In supergravity we can certainly look for
conformal deformations, so we should be able to do it directly in
the field theory as well. Actually, from the previous sections we
know that all we need to know is the action of the
super-conformal charges on the operators of the theory, and
whereas little is known about the correlation functions, the
spectrum of protected operators {\it is} known (from supergravity
for instance). {\it Roughly} the spectrum is
generated\footnote{though not freely} by bosonic fields $\phi_i$
in the ${\bf 8_v}$ of $SO(8)$ with dimension $\Delta_\phi=1/2$
(free boson dimension in 3d) and fermionic fields $\psi_\alpha$
in the ${\bf 8_s}$ of $SO(8)$ with dimension $\Delta_\psi=1$
(free fermion dimension in 3d).

The first step is to identify the super-marginal operators. We
limit ourselves to 3d $\cn=2$ susy deformations in order to be
able to use later the D-term constraint. Whether relaxing to
$\cn=1$ allows additional deformations is work in progress. 3d
$\cn=2$ comes with a $SO(2)_R=U(1)_R$ global symmetry and
introduces a decomposition of the global group $SO(8) \to SU(4)
\times U(1)$. There are two massless scalar representations on
$AdS_4 \times S^7$ - a ${\bf 840_c}=[2,0,2,0] \mbox{ of } SO(8)$
from a mode of the 3-form potential and a ${\bf 1386}=[6,0,0,0]
\mbox{ of } SO(8)$ from a mode of the metric (mixed with the warp
factor)\footnote{See \cite{S7spectrum}, but some shifts in the
conventions for $m^2$ must be performed \cite{AdSCFTS7}.}. In the
field theory they are roughly $\psi^2\, \phi^2$ and $\phi^6$
\cite{AdSCFTS7}, respectively. Since the super-marginal modes
satisfy a first order differential equation (susy being the
``square root'' of the second order equations of motion), only the
3-form modes may be super-marginal (the mode $[6,0,0,0]$ is a KK
mode for a scalar field and there is no natural first order
equation for a scalar). So we proceed to decompose the ${\bf
840_c}$ according to $SO(8) \to SU(4) \times U(1)$ (embedded such
that ${\bf 8_c} \to {\bf 6}_0 + {\bf 1}_2 + {\bf 1}_{-2}$).
Rather than analyze the differential equation we notice that
since we want the $U(1)_R$ to be preserved we need consider only
representations with zero $R$-charge, which are $({\bf
35}_\IC+{\bf 84}+{\bf 45}_\IC+{\bf 20'}
)_0=([4,0,0]_\IC+[2,0,2]+[2,1,0]_\IC+[0,2,0])_0$ where each
complex representation is accompanied by its conjugate. We expect
a conformal superpotential to have the form $W \sim h_{ijkl}\,
\phi^4$ which in 3d should have dimension 2. Hence I expect that
the susy variation equations will select the ${\bf 35}_0$, the
fourth rank symmetric tensor of $SU(4)$ together with its complex
conjugate (showing that is work in progress).

The analysis above is valid for $N = \infty$. For finite $N$ we
can take 4d $\cn=4$ as a guiding example, and expect that the
same operators will be super-marginal for some large enough $N$
(in 4d $\cn=4$ we need $N \ge 3$ for the existence of the
$d^{ABC}$ invariant of $SU(N)$), and in addition other copies of
the ${\bf 35}$ may exist by analogy with multi-trace operators
(which were impossible to make out of the $\phi^3$ deformation of
$W$ in 4d).

Now we may add the D-term to the global $SU(4)$ and conclude that
\be
 \Mc (3d,\, \cn=8) \simeq {\bf 35}/SL(4,\IC) \label{mc3} \ee
where as before the equality is local in the neighborhood of the
origin (the original membrane theory), $\cn=2$ susy is assumed and
the possible finite $N$ effects discussed above should be borne in
mind.

One may wonder whether a generalization of (\ref{mc4},\ref{mc3})
could be interesting, namely $S^N(F)/SL(N,\IC)$ where $S^N(F)$
stands here for the $N^{\mbox{th}}$ symmetric product of the
fundamental representation of $SU(N)$.

\vspace{0.5cm} \noindent {\bf Acknowledgements}

\noindent Special thanks to Ofer Aharony for an extensive
collaboration on a related problem and for comments on the
manuscript.
 I would like to thank B.
Acharya, J. Maldacena, N. Seiberg, S. Yankielowicz and E. Witten
for important discussions. I thank the Hebrew University in
Jerusalem, the Weizmann institute and Stanford University for
hospitality during the course of this work.

Work supported by DOE under grant no. DE-FG02-90ER40542, and by a
Raymond and Beverly Sackler Fellowship.


\end{document}